\documentclass[conference]{IEEEtran}
\usepackage{spconf,amsmath,graphicx}
\usepackage{gensymb}
\usepackage{hyperref}
\usepackage{float}

\title{CochCeps-Augment: A Novel Self-Supervised Contrastive Learning Using Cochlear Cepstrum-Based Masking for Speech Emotion Recognition}

\name{Ioannis Ziogas$^1$, Hessa Alfalahi$^1$, Ahsan H. Khandoker$^1$, Leontios J. Hadjileontiadis$^{1,2}$ \thanks{This work is supported by Khalifa University of Science and Technology. © 2024 IEEE. Personal use of this material is permitted. Permission from IEEE must be obtained for all other uses, in any current or future media, including reprinting/republishing this material for advertising or promotional purposes, creating new collective works, for resale or redistribution to servers or lists, or reuse of any copyrighted component of this work in other works.}} 
\address{\small $^1$Department of Biomedical Engineering, Khalifa University of Science and Technology, Abu Dhabi, UAE \\
\small $^2$Department of Electrical and Computer Engineering, Aristotle University of Thessaloniki, Greece}

\begin{document}

\IEEEoverridecommandlockouts
\IEEEpubid{\makebox[\columnwidth]{978-1-5386-5541-2/18/\$31.00~\copyright2024 IEEE \hfill}
\hspace{\columnsep}\makebox[\columnwidth]{ }}

\maketitle

\IEEEpubidadjcol

\begin{abstract}
Self-supervised learning (SSL) for automated speech recognition in terms of its emotional content, can be heavily degraded by the presence noise, affecting the efficiency of modeling the intricate temporal and spectral informative structures of speech. Recently, SSL on large speech datasets, as well as new audio-specific SSL proxy tasks, such as, temporal and frequency masking, have emerged, yielding superior performance compared to classic approaches drawn from the image augmentation domain. Our proposed contribution builds upon this successful paradigm by introducing \textit{CochCeps-Augment}, a novel bio-inspired masking augmentation task for self-supervised contrastive learning of speech representations. Specifically, we utilize the newly introduced bio-inspired cochlear cepstrogram (CCGRAM) to derive noise robust representations of input speech, that are then further refined through a self-supervised learning scheme. The latter employs SimCLR to generate contrastive views of a CCGRAM through masking of its angle and quefrency dimensions. Our experimental approach and validations on the emotion recognition K-EmoCon benchmark dataset, for the first time via a speaker-independent approach, features unsupervised pre-training, linear probing and fine-tuning. Our results potentiate \textit{CochCeps-Augment} to serve as a standard tool in speech emotion recognition analysis, showing the added value of incorporating bio-inspired masking as an informative augmentation task for self-supervision. Our code for implementing \textit{CochCeps-Augment} will be made available at: \url{https://github.com/GiannisZgs/CochCepsAugment}.
\end{abstract}

\section{Introduction}
\label{sec:intro}
\small
Retrieving information implicitly from spoken language and natural sounds has been believed to constitute one of the core learning mechanisms during development in humans, who essentially form a perception of their acoustic surroundings based on salient acoustic representations \cite{balestriero2023}. Similarly, advances in neural networks have formed the yet nascent, though promising, Self-Supervised representation Learning (SSL) field, in an effort to mimic human acquisition of knowledge that does not rely on explicitly taught or labeled paradigms \cite{balestriero2023}. 
Contrastive learning, a sub-category of SSL, leverages simple augmentations to generate multiple views of a sample, in order to foster invariance, and then encourage similarity between their feature representations \cite{balestriero2023}. 

In the area of speech and audio processing, SSL provides a compelling paradigm that can mitigate the lack of sufficiently large and well-annotated datasets \cite{chong2023masked} 
, hence facilitating the learning of intelligible speech features that enhance the performance on a multitude of downstream tasks.  SSL concepts from computer vision and natural language processing such as masking \cite{he2022MAE, devlin2018BERT} have been extended to audio processing, and have successfully exploited vast unlabeled speech datasets to create large, pre-trained-on-speech models such as, wav2vec2.0 \cite{Baevski2020Wav2vecRepresentations} and HuBERT \cite{Hsu2021HuBERT:Units}. However, SSL for audio has limitations that relate to the temporal structure of time series, as well as the heavy noise contamination usually present in audio \cite{wang2020b}, factors that limit the application of classic contrastive image augmentations to audio or audio-derived representations, such as, spectrograms and MFCCs. To that end, audio-tailored augmentation methods that are based on masking have been designed, such as, SpecAugment \cite{park2019SpecAugment} and Mask Spec \cite{chong2023masked} and have been successfully applied for contrastive learning in audio data \cite{soni2022contrastive}.

Among the recently evolving tasks are speech emotion recognition (SER); which currently plays a huge role in Human-Computer Interaction (HCI) \cite{cowie2001emotion}, as well as various healthcare \cite{dhuheir2021emotion} and e-learning paradigms in education applications \cite{li2007speech}. SER constitutes a sub-field of automatic speech recognition that benefits from feature representations of the raw audio such as spectrograms or MFCCs \cite{zou2022,he2023}. SSL masking-based approaches in particular, have proven to be very effective in the SER paradigm, yet large pre-trained models on generic speech datasets are preferred, as these representations have been proven to be beneficial for the task of SER \cite{kakouros2023}.

\begin{figure*}[ht!]
    \centering
    \includegraphics[width=\textwidth, height = 5cm]{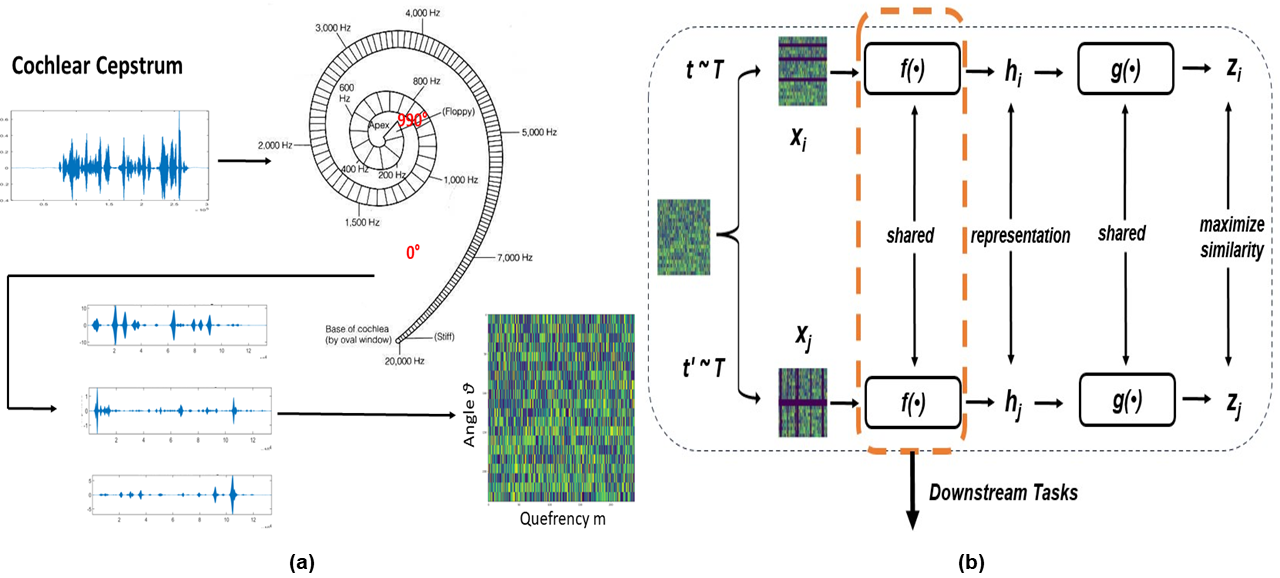}
    \caption[]{\footnotesize{\textsf{Our proposed bio-inspired \textit{CochCeps-Augment} SSL framework: (a) Cochlear Cepstrum, (b) SimCLR Contrastive pre-training}}}
    \label{fig1}
\end{figure*}

We believe that bio-inspired signal representation methods could guide contrastive learning methods to learn noise-robust intelligible features, thereby enhancing the generalization ability and efficacy of the model for the problem at hand (i.e., SER). To the best of our knowledge, this is the first work that attempts to enhance synergies between machine intelligence and human perception by adopting a bio-inspired cochlear cepstral representation of speech signals for a novel SSL framework in SER. As opposed to the well-known Mel Frequency Cepstral Coefficients (MFCCs) \cite{singh2012mfcc} and GammaTone filterbank Cepstral Coefficients (GTCCs) \cite{valero2012gammatone}, we adopt the Cochlear Filterbank Cepstral Coefficients (CFCCs), also referred to as cochlear cepstrogram (CCGRAM), that mimic both the function and the structure of the human cochlea \cite{alfalahi2024spiral}. The human cochlea is characterized by a spiral structure that encodes the frequency-position map; which is the essence of the superb frequency resolution of the human ear. In particular, the cochlear spiral is geometrically composed of a bit more than two and a half turns spanning $\theta =0\degree$ at the base (high frequency hearing, up to 20 kHz) of the cochlea to $\theta = 990\degree$ at the apex (low frequency hearing, up to 10 Hz) of the cochlea \cite{rask2012human}. 


In light of the aforementioned, the contribution of our work is a novel augmentation method which we call \textit{CochCeps-Augment}. Our method draws inspiration from SpecAugment \cite{park2019SpecAugment}, however, in contrast to SpecAugment, our method operates on the image representation of the CCGRAM of the input audio by applying masking along the angle and quefrency axis, therefore encouraging self-supervision to attend to meaningful tonotopically-organized audio properties that are intelligible to humans. \textit{CochCeps-Augment} is simple and cost-effective to be applied during self-supervised pre-training and due to the bio-inspired nature of the CFCCs \cite{alfalahi2024spiral}, exhibits enhanced noise robustness which is highly desirable in audio analysis. The results of applying \textit{CochCeps-Augment} on the K-EmoCon dataset \cite{Park2020K-EmoConConversations} for the first time in a speaker-independent manner, via a self-supervised contrastive learning scheme, showcase the potentiality of our bio-inspired masking augmentation task. We believe that \textit{CochCeps-Augment} will not only enhance SER tasks, but will also likely unlock new avenues for various applications in speech and acoustic signal processing; especially by blending human auditory mechanisms into SSL. An overview of our method is presented in Figure \ref{fig1}.

\section{Cochlear Cepstrum Background}
The human ear shows remarkable ability in recognizing speech content under variable conditions of background noise, thereby inspiring a range of signal processing and representation methods \cite{alfalahi2023cochlear,alfalahi2023cochlear2}. We employed a recently proposed bio-inspired, noise-robust feature space, called Cochlear Filterbank Cepstral Coefficients (CFCCs) \cite{alfalahi2024spiral}. Such CFCCs replicate both the structure and the function of the human spiral cochlea and are based on the concept of the cochlear transform (CT) \cite{alfalahi2023cochlear2}. The latter is briefly a novel and general signal processing framework that mimics the active and non-linear multiscale analysis of the cochlea for acoustic signals. The resulting \textit{Cochlear Modes} are transformed to the cochlear cepstral space by means of logarithmic transformation and Discrete Cosine Transform (DCT) of the resulting modes' energy. Hence, for the computation of the CFCCs of a signal $p(t)$, we start by computing the orthogonal cochlear modes (in the frequency domain), denoted $FCT$ that result from the CT, as follows: 

\begin{equation}
\centering
 FCT_p(\theta,\omega) = \sqrt{\theta} P(\omega) \Phi^{*}(\theta, \omega ), 
\label{Eq1}
\end{equation}
\noindent 
where $\theta$ is the angle along the spiral cochlear space; namely $\theta \in [0\degree, 990\degree]$. It should be noted that the tonotopic place-pitch map is defined as follows: 
\begin{equation}
f(\theta)=165.4(3251^{2.1}(\theta + 177.3)^{-2.1 \times 1.149} - 0.88), 
\label{Eq2}
\end{equation} 
\noindent 
where $f$ is the frequency and $\theta$ is the angular position across the spiral cochlea. It should be noted that $\theta =0\degree$ corresponds to the base (High frequency region) and $\theta =990\degree$ corresponds to the apex of the cochlea (Low frequency region). Afterwards, we compute the log magnitude spectrum and the DCT to extract the tonotopically (spatially) organized cochlear cepstral coefficients. By definition, the real CFCCs at a specific angular position $\theta$, is: 
\begin{equation} 
\centering 
CFCC(m, \theta) = \sqrt{\frac{2}{K}} \sum_{k=1}^{K} log(X_{k}) cos [\frac{\pi k}{K}(m-\frac{1}{2})], 
\end{equation}
\noindent 
where $CFCC(m, \theta)$ is the mth cochlear cepstral coefficient at angle $\theta$ and $m$ is the cochlear quefrency index, $1 \leq m \leq M$, $K$ is the number of cochlear modes and $X_{k}$ is the energy of the kth cochlear mode (i.e., $FCT_{p}(\theta)$).

\section{Proposed Method}
\subsection{Cepstral Masking Augmentation Policy}
To enhance the expressivity of the CCGRAM features in modelling speech in a SSL setting, we design our masking augmentations to operate along both the angle $\theta$ and quefrency axis $m$. In doing so, we encourage self-supervision to infer the masked frequency tones arising from particular angular positions along the cochlea and masked speech segments along the time domain, both individually and in a combined fashion. Therefore, in a given sample with two cepstral-augmented views, the model should be able to predict the missing tones given the presence of other, contextually and acoustically relevant tones. Hence, it becomes evident that the proposed \textit{CochCeps-Augment} proxy task, is relevant both for generic speech and audio processing, as well as in the context of SER. This leads us to the following three augmentation transforms:

\begin{enumerate}
    \item \textbf{Angle masking}, applied along the angle $\theta$ axis, so that $\Phi$ distinct masks are applied. Each mask spans $\phi$ consecutive angle bands, i.e. $[\theta_0,\theta_0 + \phi)$, where $\phi$ is sampled randomly from a uniform distribution from 0 to $\Phi$, and $\theta_0$ is a random angle sampled from $[0, 990 - \phi)$. 

    \item \textbf{Quefrency masking}, applied along the quefrency indexes $m$, so that $Q$ distinct masks are applied. Each mask spans $Q$ consecutive quefrency indexes, i.e. $[m_0,m_0 + q)$, where q is sampled randomly from a uniform distibution from 0 to $Q$, and $m_0$ is chosen from the $[0,M - m)$.

    \item \textbf{Cepstral masking}, defined as the simultaneous masking of both angle bands and quefrency indexes according to the above parameters.
\end{enumerate}

\begin{figure}[t!]
    \centering
    \includegraphics[width=1\linewidth]{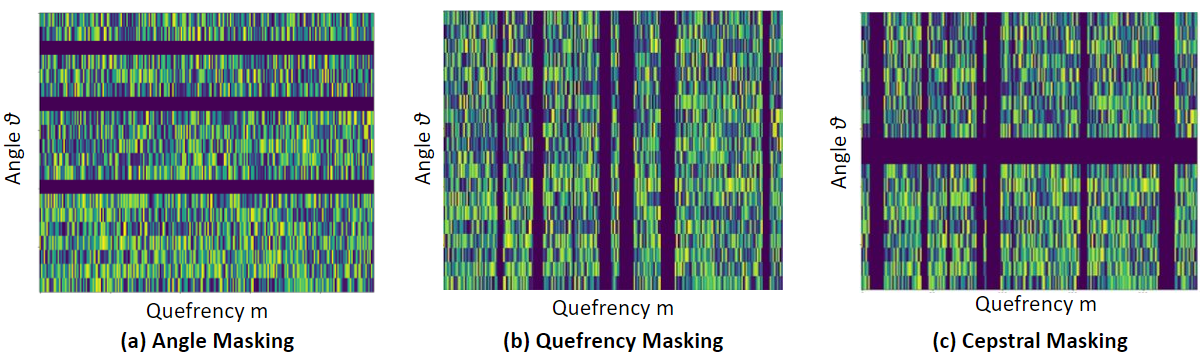}
    \caption[]{\footnotesize{\textsf{Family of the proposed CCGRAM augmentations applied on a single sample (for $\theta \in[0\degree,990\degree]$ and quefrency index $m$).}}}
    
    \label{fig:ceps_augmentations}
\end{figure}

Figure \ref{fig:ceps_augmentations} shows an example of the application of our proposed augmentations to an input CCGRAM.

\subsection{SimCLR - Self-Supervised Contrastive Learning of Representations}

To take advantage of the masking benefits that \textit{CochCeps-Augment} adds to the learning process of a SSL system, we adopt SimCLR as our pre-training framework \cite{chenT2020}. In SimCLR, two augmentation transforms ($t \sim T$ and $t' \sim T$) are sampled from the family of the proposed \textit{CochCeps-Augment} transforms $T$. SimCLR promotes similarity between two augmented views of an input sample, designed to maintain invariance on information axes that are deemed redundant. Specifically, meaningful representations can be learned by maximizing the similarity between positive pairs (views of the same sample) and minimizing the similarity between negative pairs (views belonging to different samples) in a latent space. To that end, every sample CCGRAM $x$ is masked twice in order to produce two distinct augmented views, $x_i$ and $x_j$. By forward-propagating the two views through a $shared$ $encoder$ $f(.)$, two representations $h_i$ and $h_j$ are obtained, which are further projected through a $projector$ $g(.)$ to obtain the final representations $\mathbf{z}_i$ and $\mathbf{z}_j$. 

The self-supervision without labels in SimCLR, is facilitated by the NT-Xent (normalized temperature-scaled cross-entropy) loss function, which is given by:

\begin{equation*}
\mathcal{L}_{\text{SimCLR}}(i, j) = -\log\left(\frac{\exp(\text{sim}(\mathbf{z}_i, \mathbf{z}_j) / \tau)}{\sum_{k=1}^{2N}\mathbf{I}[k\neq i] \exp(\text{sim}(\mathbf{z}_i, \mathbf{z}_k) / \tau)}\right)
\end{equation*}

where $\mathbf{z}_i$ and $\mathbf{z}_j$ are the representations of augmented views $i$ and $j$, $\text{sim}(\mathbf{z}_i, \mathbf{z}_j) = \frac{\mathbf{z}_i \cdot \mathbf{z}_j}{\|\mathbf{z}_i\|\|\mathbf{z}_j\|}$ is the cosine similarity, $\tau$ is a temperature parameter, $\mathbf{I}[\cdot]$ is the indicator function and $N$ are the number of samples in a batch. 

In the pre-training phase, the encoder $f(.)$ and the projector $g(.)$ are trained end-to-end and evaluated through the NT-Xent loss. 
In downstream evaluations, the projector is discarded and the encoder $f(.)$ is used as a feature extractor that yields the features $h_i$ and $h_j$. 


\section{Experiments}
\label{sec:pagestyle}
\subsection{K-EmoCon Database}
The K-EmoCon database contains multi-modal recordings from 32 participants divided in 16 groups \cite{Park2020K-EmoConConversations}. Participants engaged in a dyadic debate in English (~10 minutes) during which multi-modal data acquisition took place. 
For emotional labeling, we adopt the self-rated arousal/valence (A/V) space quadrant scheme (see Table \ref{table1}), where the provided integer ratings in the range $[1,5]$, are binned to a specific quadrant based on their combined arousal and valence, yielding four classes: LowA/LowV (LALV), LowA/HighV (LAHV), HighA/LowV (HALV), HighA/HighV(HAHV). 

\begin{table}
\caption{Quadrant Annotations: Arousal (a) /Valence (v) values are binned to one of the four quadrants of the A/V space.}
\begin{center}
\begin{tabular}{|c|c|c|} 

\hline
 & Arousal & Valence \\
\hline
LALV & $a < 3$ & $v < 3$ \\
\hline
LAHV & $a < 3$ & $v \geq 3$ \\
\hline
HALV & $a \geq 3$ & $v < 3$ \\
\hline
HAHV & $a \geq 3$ & $v \geq 3$ \\
\hline

\end{tabular}
\end{center}
\label{table1}
\end{table}

Data collection in K-EmoCon was modelled as a natural conversational setting, hence various sources of noise are present. Moreover, the problem of source separation needs to be addressed in speech segments where participants and the referee's voices overlap, as well as when heavy degradation occurs due to loud, non-conversation-related sounds. To that end, we design a combined pre-processing strategy which is mainly focused on source separation to obtain speaker-independent speech segments. The raw audio is first downsampled from the initial sampling rate 22.5 kHz to 16kHz, followed by max scaling. Silent segments are then removed with a Short-Time Fourier Transform energy thresholding \cite{giannakopoulos2009VAD}. We manually annotate speech segments identified as sound and based on their content, we apply one of two source separation techniques. For stationary noise, we apply the WTST-NST filter \cite{hadjileontiadis2000WTST} to isolate the non-stationary speech. For speaker separation, we utilize a pre-trained on the Librimix dataset Sepformer model that outputs two speaker signals \cite{ravanelli2021}. Finally, resulting speech segments are scaled again, and segmented into 3 second windows. To ensure uniform input to our deep learning models, we zero-pad segments that are shorter than 3s, and discard segments that are shorter than 1s. 

We compute the CCGRAM of an input 3-second segment according to \cite{alfalahi2024spiral}, using Hamming windows of 25ms duration with 50\% overlap. We set the angle spacing between cochlear modes at $\theta = 45\degree$, resulting in CCGRAMs of size 20x239. 


\subsection{Self-Supervised Pre-training}
Prior to any processing, we split the data into 4 distinct folds: pre-training (26 speakers), validation (2 speakers), fine-tuning (2 speakers) and test (2 speakers). For SSL pre-training we use the pre-training and validation folds, while for evaluation we use the pre-training, validation and test folds in the linear probing, and the fine-tuning, validation and test folds in the fine-tuning. 
We follow a speaker-independent approach, that relies on a 5-fold cross-validation scheme of K-EmoCon. Specifically, we design the folds in order to ensure that all speakers are present once at each fold, and that all folds contain only whole speakers. 

For our augmentation pipeline, we first z-normalize all images in a fold with the mean value and standard deviation of that fold. Then, we generate two views of each input CCGRAM by randomly selecting one of the three masking transforms in the family of \textit{CochCeps-Augment} for each view. The parameters $\Phi$ and $Q$ for the angle and quefrency masking are evaluated as 2 and 5, respectively. After masking, we resize the CCGRAM to a size of 239x239 with nearest-neighbors interpolation. We pre-train from scratch a ResNet18 to be used as the encoder $f(.)$ \cite{he2015ResNet}, where we slightly modify the input convolutional layer to accept single-channel intensity images. The projector $g(.)$ is implemented as a two-layered linear head with an output dimension of 256. For the NT-XEnt loss, we choose a temperature parameter of 0.07. For our implementation via PyTorch, we trained on a machine with four NVIDIA RTX6000 ADA GPUs, with a batch size of 64 for 1000 epochs on each pre-train fold. The LARS optimizer was used \cite{you2017LARS}, with a starting learning rate of 0.1 and a weight decay parameter of 1e-6. A warm-up period of 10 epochs is followed by training with the cosine decay scheduler \cite{loshchilov2017}. 

\subsection{Evaluations}
We evaluate the performance of our SSL approach through two schemes: linear probing and fine-tuning. It should be noted that in the evaluations, we do not apply \textit{CochCeps-Augment} but only the z-normalization and resizing operations. Linear probing refers to the evaluation of the frozen encoder $f(.)$ through a simple linear head, that is used to perform the downstream task. Linear probing constitutes a simple evaluation method and allows us to understand the pure quality of the features that the encoder learned during the pre-training phase \cite{balestriero2023}. Moreover, to test the ability of the contrastive self-supervision to recover the CCGRAM-encoded speech information, we conduct a sanity check where the flattened CCGRAM is fed directly to the linear probe. We expect that this evaluation will reveal the information gain that our proposed pre-training approach contributes to the end result. From a technical perspective, this is attributed to the fact that flattening disrupts the spatio-temporal cochlear cepstral features. Fine-tuning on the other hand, utilizes a non-linear head for downstream classification, and refers to the classic process of fine-tuning the whole network, including the encoder. We train with linear probing on the same data that was used for the pre-training, and evaluate on the left-out test fold. For fine-tuning, we tune on the fine-tuning fold and evaluate on the test fold.  On both settings, we train with a batch size of 16, for 50 epochs to avoid overfitting. 
The Adam optimizer is used \cite{kingma2014Adam} with a starting learning rate of 1e-4 for linear probing and 5e-6 for fine-tuning, with weight decay of 1e-6 and a cosine decay schedule. For performance evaluation, we use the weighted accuracy and weighted F1-score.

\begin{table}[t!]
\caption{SSL Evaluation Results on K-EmoCon: Averages over 5-fold Speaker-Independent Cross-Validation}
\begin{center}\resizebox{.5\textwidth}{!}{
\begin{tabular}{|c|c|c|} 

\hline
 & Weighted Accuracy & Weighted F1 \\
\hline
Linear Probing – Flattening & 0.42 & 0.45 \\
\hline
Linear Probing – ResNet18 & 0.61 & 0.50 \\
\hline
Fine-Tuning – ResNet18 & \textbf{0.69} & \textbf{0.57} \\ 
\hline

\end{tabular}}
\end{center}
\label{results}
\end{table}

\section{Results and Discussion}
\label{sec:typestyle}
The results of our proposed \textit{CochCeps-Augment}-driven SSL approach for the SER task can be seen in Table \ref{results}. We employed two main evaluation schemes for the proposed approach; namely linear probing and fine-tuning. As illustrated in Table \ref{results}, for the linear probing task, the weighted accuracy/F1 score are 0.42/0.45 for the flattening sanity check, and 0.61/0.50 for the ResNet18 pre-trained encoder; respectively. However, the performance is significantly enhanced upon the fine-tuning stage; yielding an accuracy and F1 score of 0.69 and 0.57 respectively. This indicates that self-supervision through \textit{CochCeps-Augment} and further refinement of the learned features in the downstream speaker-independent SER task, can indeed uncover meaningful non-redundant representations. 

As K-EmoCon is a small-scale emotion recognition corpus, we believe that pre-training on large-scale speech corpora and validating on more emotional speech corpora, would benefit the power of our conclusions. 
Moreover, longer training times have been proven beneficial for SSL feature extractors \cite{soni2022contrastive} and thus could enhance our results. It is still not clearly understood how each distinct masking scheme of \textit{CochCeps-Augment} affects the learned representations; angle masking for instance, steers the learning towards perceived tonotopical relationships in an input utterance, whereas quefrency masking promotes learning of contextual links between speech segments. However, simultaneous or excess masking of angle and quefrency content may irreversibly degrade the CCGRAM and hence, the capacity of self-supervision to recover information from the perturbed CCGRAM. Finally, \textit{CochCeps-Augment} is a task-agnostic SSL proxy task for speech; therefore, we hypothesize that representations learned through \textit{CochCeps-Augment} could improve speech recognition systems in a variety of tasks, i.e. speaker identification, automatic speech recognition, audio events classification. Investigation towards these directions is already underways.  

\section{Conclusion}
\label{sec:majhead}
In this work, we have presented our bio-inspired masking augmentation method, namely \textit{CochCeps-Augment}, for learning self-supervised speech representations through contrastive learning in the context of SER. We demonstrated, for the first time, how \textit{CochCeps-Augment} can be seamlessly and cost-effectively integrated in a resource-demanding contrastive SSL setting through SimCLR, in the context of SER, with promising results. This novel approach showcases how human perception of sounds, encapsulated in the signal processing framework of the cochlea, can be placed in the epicentre of a speech self-supervision model, which by design tries to implicitly mimic the way humans perceive auditory events. 

\vfill\pagebreak



{\footnotesize
\bibliography{strings,refs, references_giannis}}

\end{document}